# PRAGMATIC INFORMATION AND GAIAN DEVELOPMENT --- FIRST THOUGHTS

Edward D. Weinberger

Aloha Enterprises, Inc.

370 Central Park West, Suite 110

New York, NY 10025

edw@panix.com

> *"In the Beginning, God created the Heavens and the Earth."*
>
> -- Genesis 1:1
>
> *"There is grandeur in this view of life, with its several powers, having been originally breathed by the Creator into a few forms or into one; and that, whilst this planet has gone cycling on according to the fixed law of gravity, from so simple a beginning endless forms most beautiful and most wonderful have been, and are being, evolved."*
>
> -- Last sentence, Darwin's *Origin of Species*

## Abstract

*The scientific community believes in the theory of evolution with a passion that rivals that of any religious belief. This passion extends beyond the irrefutable evidence of the fossil record to the familiar claims of "survival of the fittest" and random mutation. Yet the theory of natural selection has, to the knowledge of the present author, never been tested against the alternative hypothesis that evolution is, in fact, the ongoing development of a single, world-spanning super-organism. Just what kind of evidence would settle this question is not clear, so this paper is intended to begin the discussion. It does so by suggesting how a new, quantitative theory of "pragmatic information," first presented in Weinberger (2002), might detect the widespread temporal and inter-species connections that a developmental view of evolution would imply.*

## Introduction

The phrase "the theory of evolution" belies the *fact* of evolution. However one quibbles with any particular interpretation of the fossil record, the mere existence of fossils that differ dramatically from anything currently living makes clear that some kind of evolution must have taken place.

Less obvious is the cause of evolution, and it is here that reasonable doubts can be raised. As is well known, the current orthodoxy of natural selection makes two distinct claims:

*i )* Every biological population has a variety of phenotypes, some of which make their representatives better adapted to their environment. As a result, these phenotypes are more likely to contribute representatives to the next generation.

*ii )* The primary source of biological variation is random mutation of the underlying genotype.

No doubt, biological populations do undergo natural selection, if only because "survival of the fittest" is a tautology. What else can fitness mean besides the expected contribution of progeny to future generations? The question is therefore not whether natural selection exists, but to what degree is it sufficient to account for Darwin's "endless forms most beautiful and wonderful… [that] are being evolved." Natural selection has been seen repeatedly in the increasing resistance of bacteria to antibiotics or the artificial selection imposed by human animal breeders. However, in the latter case, breeding for certain desirable hereditary traits, as dog breeders have been doing for centuries, can also lead to a decline in other desirable traits (Grandin, 2005). The conflicts inherent in optimizing for multiple traits can easily lead to locally, but not globally optimal genotypes/phenotypes. The possibility, even the likelihood, of local fitness optima that might frustrate evolutionary "hill-climbing" is explored Kauffman (1993).

We also know that there are other forces at work besides natural selection. For example, Kimura (1990) proposes a theory of neutral drift, based on the idea that multiple genotypes can produce phenotypes of equal fitness. An early classic of theoretical biology, Thompson's *On Growth and Form* (Thompson, 1992), observes that selection must take place within the non-trivial constraints imposed by the laws of physics and chemistry. A host of more recent authors, beginning with Turing (1952), have demonstrated the importance of the reaction-diffusion equation of physical chemistry in producing a wide variety of morphologies. Finally, Mayr (2001) stresses the importance of geographical isolation in the emergence of new species.

Natural selection's second claim, that the variation required for the action of natural selection is generated by "random" mutation, is somewhat more problematic because randomness can never be verified directly. Randomness is generally defined as something that is incapable of being predicted from *any* algorithm, so, in principle, a demonstration of randomness requires failed predictions by all possible algorithms.

A second objection to the "random mutation" hypothesis is the unlikelihood of "order from disorder." A chestnut of statistical mechanics is the idea that, given enough time and typing paper, a monkey pecking at a typewriter randomly will type the entire text of Shakespeare's *Hamlet*. What is left out of this story is just how long the wait will be. As a studious undergraduate, I estimated that the waiting time for a diligent (and very long lived!) monkey to produce such a masterpiece with probability one-half was $10^{190,000}$ years, rendering this outcome effectively impossible. The point of this calculation is the extremely long waiting time until this event transpires. Perhaps "endless forms … are being evolved," but the random process of mutation and natural selection may simply be incapable of producing these forms in the long, but finite time available.

It is therefore reasonable to ask

*i )* can *any* of the above, either by themselves or in combination, really account for the complexity of living things and the unity of the biosphere?

*ii )* how much does natural selection, in particular, account for this complexity?

If we set aside the baggage that all too often accompanies such claims, we are forced to admit that the possibility of some kind of overall plan of evolution is at least conceivable. Ever since Adam Smith introduced the idea of the "invisible hand" into economics, students of that discipline have held "market forces" in high regard. Surely this is precisely the kind of supra-rationality postulated for the Intelligent Designer[1]. Or imagine the situation of a single "conscious" embryonic cell that is trying to make sense of "evolutionary" change in the developing organism, some of which is known (by humans) to be mediated by "survival of the fittest." We humans use the word "development" to describe the process precisely because it seems to follow a specific, pre-determined sequence of steps, in contrast to natural selection, which is believed to be the result of random genetic mutation.

However, this may or may not be what our hypothetical cell would observe. It is therefore reasonable to ask a final question,

*iii )* is there an experiment, or set of experiments that could detect whether the ongoing evolution of biological species is, in fact, the development of an "ur-organism"?

This idea is not that big a leap beyond the Gaia hypothesis of J. E. Lovelock (1979), who shows that living things are responsible for maintaining a variety of planet-wide homeostatic equilibria[2].

---

[1] A testament to the collective intelligence of markets is the ongoing success of the University of Iowa futures market for the outcome of elections, http://www.biz.uiowa.edu/iem/. Describing itself as " [a] real-money futures market in which contract payoffs depend on economic and political events such as elections", its futures prices have been better predictors of recent presidential elections than any opinion poll available to the public.

[2] One of the more remarkable of these is that of temperature, because, according to Lovelock, the sun is known to have increased in brightness by 25% since the life began on Earth.

Accordingly, we choose the phrase "Gaian development" as a name for the view of evolution described in *iii*.

So what light might a theory of information shed on all of this? Development contrasts with the orthodox view of evolution in that the former is exquisitely coordinated and the latter is haphazard. Thus, a significant piece of evidence for Gaian development would be the detection of planet-wide, causally connected biospheric changes[3], as well as evidence that causal connectivity also extended throughout the entire history of life on earth. This connectivity must be mediated by some kind of information transfer, so perhaps a study of this transfer might hold clues to the questions raised above.

Weaver's lucid introduction to the paper by Shannon that began the study of information theory (Shannon and Weaver, 1962) observes that the effectiveness of a communications process can be measured by answering any of the following three questions:

A. *How accurately can the symbols that encode the message be transmitted ("the technical problem")?*
B. *How precisely do the transmitted symbols convey the desired meaning ("the semantics problem")?*
C. *How effective is the received message in changing conduct ("the effectiveness problem")?*

Weaver then notes that Shannon's paper --- and thus the entire edifice of what is now known as "information theory" --- concerns itself only with the answer to question A. Eigen (1971) makes clear the need for a theory of information semantics to advance evolutionary theory by writing

> *This complementarity between information and entropy shows clearly the limited application of classical information theory to problems of evolution…. It is of little help as long as information has not yet reached its "full meaning", or as long as there are still many choices for generating new information. Here we need a new variable, a "value" parameter, which characterizes the level of evolution.*

The various theory of "complexity" that have been proposed to address question B ignore the fact that a signal gets parsed differently depending on how the information in it is to be acted upon. The meaning of a piece of information can only be assessed by its impact, by the change in behavior of the receiver. In other words, question B can only be addressed by addressing question C, which is what the theory of pragmatic information attempts to do.

The essence of this theory, first outlined in Weinberger (2002), is that Shannon's communications theory can indeed be extended to address question C. The responses of question C can be viewed as the ultimate end of the communication of question A, and thus amenable to similar mathematical analysis. Weinberger (2002) also shows that the pragmatic information of the quasi-species equation is always increasing, thus providing the value parameter that Eigen requires.

---

[3] In fact, the fossil record indicates such changes, in the form of various mass extinctions, but the temporal coordination required for Gaian development has yet to be found, at least to the present' author's knowledge.

Because no single paper can do justice to these questions, we present here only a few initial thoughts on the subject, in the hope of simulating further discussion. After presenting a summary of the theory of pragmatic information, we indicate how it might help to detect the connections between species required by Gaian development. Detecting connections through time requires an extension of the theory to include a computational model of the receiver. Since this is work in progress, we only sketch the relevant theory and how it might be applied. We conclude with a summary, a few thoughts about future research, and the distinction between Gaian development and the pseudo-theory of "Intelligent Design".

## *An Introduction to the Theory of Pragmatic Information*

### *The Definition*

This section is a summary of the theory as presented in Weinberger (2002), which, to the best of our knowledge, is where the present theory was first proposed. Because of the novelty of the theory, much of that paper was devoted to establishing that the definition of pragmatic information given there was both unique and intuitively reasonable.

The formal definition of pragmatic information is informed by the diagram shown in Figure 1. This diagram is to be interpreted as follows: a decision maker, $\mathcal{D}$, in some currents state, $s$, must choose among a set $\{a_1, a_2, \ldots a_M\}$ of alternatives, assumed to be finite only for simplicity of exposition. These choices lead to outcomes $\omega \in \Omega = \{\omega_1, \omega_2, \ldots, \omega_N\}$, some or all of which the decision maker prefers to others. However, $\mathcal{D}$'s ability to produce a desired outcome is limited, perhaps because of one or more of the following:

- $\mathcal{D}$ has imperfect information about which of the $\omega$'s is, in fact, the best,
- $\mathcal{D}$ is unable to process the available information optimally,
- $\mathcal{D}$ is unable to guarantee that the decision made is the one actually implemented, possibly because of "environmental noise", $e$,
- the mapping from the $a$'s to the $\omega$'s is not deterministic,

We reflect this indeterminacy by assuming that $\mathcal{D}$'s decision would result in a selection from the alternatives with respective probabilities $\boldsymbol{q} = (q_1, q_2, \ldots q_M)$. $\mathcal{D}$ is then presented with a "message" $m$, one of the possible messages in message ensemble $\mathcal{M}$. Even with this new data, the best that $\mathcal{D}$ can do is to update the respective selection probabilities to $p_m = (p_{1|m}, p_{2|m}, ..., p_{N|m})$ in such a way that the probability of a superior outcome is increased. This framework suggests the following

**Definition.** The ***pragmatic information***, $I_\mathcal{M}(p; q)$, of the message ensemble $\mathcal{M}$ is the information gain in going from $q$ to $p_m$, averaged over all messages $m \in \mathcal{M}$, i.e.

$$I_M(p;q) = \sum_{i,m} p_{i|m} \varphi_m \log_2 \left( \frac{p_{i|m}}{q_i} \right)$$

$$= \sum_{i,m} p_{i,m} \log_2 \left( \frac{p_{i,m}}{\varphi_m q_i} \right) \quad (1)$$

where $\varphi_m$ is the marginal probability that message $m$ was sent and $p_{i,m}$ is the joint probability that message $m$ was sent and outcome $\omega_i$ was realized.

It will be assumed that, included in $\mathcal{M}$ is the "empty message", $\nu$, which, by definition, results in no change to the *a posteriori* probabilities. The presence or absence of the empty message in $\mathcal{M}$ makes no difference to the computation, as the terms involving $\nu$ evaluate to zero. However, the ability to write $q_i = p_{i|\nu}$ will eliminate potential confusion in the discussion of joint and conditional pragmatic information below.

(Figure 1 about here…)

Figure 1 has some notable additions to similar diagrams included in discussions of "classical" information theory, beginning with Shannon and Weaver (1962). This is because standard information theory considers only the receiver's ability to distinguish the message from the other messages that could have been sent. Here, however, the explicit inclusion of a decision maker is central to the theory[4].

*Some Basic Lemmas*

We will be using the following, which are also presented for their intrinsic interest, in the sequel. Note that the numbering of the lemmas here are not those of Weinberger (2002).

---

[4] See Weinberger (2002) for additional discussion.

**Lemma 1:** If the symbols $\mathcal{M}$ and $\Omega$ have the same meaning as above, pragmatic information has the following three equivalent characterizations in terms of the (Shannon) entropy (Shannon and Weaver, 1962).

Let
$$\mathcal{H}(\mathcal{M}) = - \sum_m \varphi_m \log_2 \varphi_m,$$

the entropy of the message ensemble $\mathcal{M}$, let
$$\mathcal{H}(\Omega) = - \sum_i q_i \log_2 q_i,$$

the entropy of the outcome ensemble O, let
$$\mathcal{H}(\Omega \mid \mathcal{M}) = - \sum_{i,m} p_{i,m} \log_2 p_{i|m},$$

the conditional entropy of the outcomes, given the various messages, and let
$$\mathcal{H}(\mathcal{M} \mid \Omega) = - \sum_{i,m} p_{i,m} \log_2 p_{m|i},$$

the conditional entropy of the messages, given a particular outcome. Then
$$\mathcal{I}_\mathcal{M}(p; q) = \mathcal{H}(\mathcal{M}) + \mathcal{H}(\Omega) - \mathcal{H}(\Omega, \mathcal{M})$$
$$= \mathcal{H}(\Omega) - \mathcal{H}(\Omega \mid \mathcal{M})$$
$$= \mathcal{H}(\mathcal{M}) - \mathcal{H}(\mathcal{M} \mid \Omega)$$

Because $\mathcal{H}(\mathcal{M} \mid \Omega) \geq 0$, it follows immediately from the third identity that

**Lemma 2:** The pragmatic information of a message ensemble is always bounded above by its Shannon entropy.

We also need the notions of joint and conditional pragmatic information. The joint pragmatic information of message ensembles $\mathcal{M}$ and $\mathcal{N}$, $\mathcal{I}_{\mathcal{M},\mathcal{N}}(p; q)$, is where $p_{\omega,m,n}$ is the joint probability that both $m$ and $n$ were sent and $\omega$ was realized and $p_{\omega|m,n}$ is the

$$I(p; q) = \sum_{\omega, m, n} p_{\omega, m, n} \log_2 \left( \frac{p_{\omega|m,n}}{q_\omega} \right),$$

conditional probability that $\omega$ was realized, conditional on both $m$ and $n$ being sent.

Finally, we have the all-important

**Lemma 3:** $\qquad \mathcal{I}_{\mathcal{M},\mathcal{N}}(p; q) = \mathcal{I}_{\mathcal{M}|\mathcal{N}}(p; q) + \mathcal{I}_\mathcal{N}(p; q)$

*Proof:*

$$\begin{aligned}
\mathbf{I}_{M,N}(p;q) &= \sum_{i,m,n} p_{i,m,n} \log_2 \left( \frac{p_{i|m,n}}{p_{i|n}} \frac{p_{i|n}}{q_i} \right) \\
&= \sum_{i,m,n} p_{i,m,n} \log_2 \left( \frac{p_{i|m,n}}{p_{i|n}} \right) + \sum_{i,m,n} p_{i,m,n} \log_2 \left( \frac{p_{i|n}}{q_i} \right) \\
&= \sum_{i,m,n} p_{i,m,n} \log_2 \left( \frac{p_{i|m,n}}{p_{i|n}} \right) + \sum_{i,n} p_{i,n} \log_2 \left( \frac{p_{i|n}}{q_i} \right) \\
&= \mathbf{I}_{M|N}(q) + \mathbf{I}_N(q)
\end{aligned}$$

The pragmatic independence of two message ensembles, $M$ and $N$, is the condition that

$$\mathbf{I}_{M|N}(p;q) = \mathbf{I}_M(p;q).$$

We note in passing that this condition is *not* equivalent to the independence of the individual message probabilities, expressed mathematically as $p_{m,n} = p_m p_n$. Once again, Weinberger (2002) has details.

*The Pragmatic Information of a Process*

The remainder of this paper will be concerned with the pragmatic information of various processes. For the purposes of the applications below, only cases involving one or two points in time need be considered. For example, we might want to make predictions at time $t$ that use the state of the

$$\sum_{i,k} p_{i,k}(t_1,t_2) \log_2 \left[ \frac{p_{i,k}(t_1,t_2)}{p_i(t_1) p_i(t_1)} \right],$$

process at $t_1 < t$ to predict whether the process assumes the same states at at $t < t_2$. In the general case, where the joint probability of observing the process in state $i$ at time $t_1$ and state $j$ at time $t_2$ is $p_{i,k}(t_1, t_2)$, the pragmatic information is

In the limit of complete novelty (maximum Shannon entropy), $p_i(t_1) = p_k(t_1) = p_k(t_2)$, for all states $i$ and $k$, and for all times $t_1$ and $t_2$. Since we also have $p_{i,k}(t_1, t_2) = p_i(t_1) p_k(t_2)$ in that case (Otherwise, the conditional entropy of the state at $t_2$, given the state at $t_1$ will not be a maximum.), it follows that

**Lemma 4:** The pragmatic information of a process is zero in the limit of complete novelty.

We note the important role that additivity plays in the above. We also note that this result, combined with Lemma 2, above, proves a conjecture of Atmanspacher and Scheingraber (1991).

*An Application to Evolutionary Dynamics*

In this section, we apply the above theory to an evolving population of biological "replicators." Each individual within this population belongs to one of $N$ well defined sub-populations, each of which can be assumed to be in the large population (continuous) limit. Also, each individual has some sub-population dependent reproduction rate, which may also depend on the magnitude of other sub-populations. We identify the probability vector $\boldsymbol{p} = (p_1, p_2, \ldots p_i, \ldots)$ with the relative frequencies of the replicator sub-populations. This vector can also be interpreted as the probability of choosing a single individual at random from the overall population and finding that this individual belongs to each of the successive sub-populations. Because the samples at successive times $t_i$ and $t_k$ are independent, the pragmatic information in this case is given by

$$\sum_k p_k(t_2) \log_2[p_k(t_2)/p_k(t_1)].$$

Weinberger (2002) demonstrates that one of the simplest of evolutionary dynamical systems, the quasispecies, given by

$$d\boldsymbol{p}(t)/dt = \mathbf{W}\,\boldsymbol{p}(t) - \boldsymbol{p}(t)\left[\mathbf{1}^T \mathbf{W}\,\boldsymbol{p}(t)\right] = \left[\mathbf{W} - \Omega(t)\right]\boldsymbol{p}(t),$$

with $\mathbf{1}$ a column vector of all 1's, $\mathbf{W}$ a square matrix whose diagonal elements $W_{ii}$ represent the (constant) rates at which replicators of type $i$ are copied successfully, and whose off-diagonal elements, $W_{ij}$, represent the (constant) rates at which type $j$ is produced because of mutations while copying type $i$, $\left[\mathbf{1}^T \mathbf{W}\,\boldsymbol{p}(t)\right] = \Omega(t)$, and $^T$ is the transpose operation. Both this equation and the more general systems below satisfy $\mathbf{1}^T \boldsymbol{p}(t) = 1$ for all $t$.

We might try to generalize this result to the so-called replicator dynamics discussed in detail in Hofbauer and Sigmund (1988), given by

$$d\boldsymbol{p}/dt = \mathrm{diag}\left[\boldsymbol{p}\right]\left[\boldsymbol{f}(\boldsymbol{p}) - \mathbf{1} <f>\right].$$

Here, diag[$\boldsymbol{p}$] is the diagonal matrix with the components of $\boldsymbol{p}$ along its main diagonal, $\boldsymbol{f}$ is a vector valued "fitness" function that maps $R^N$ into the unit simplex, and $<f>$ is $\mathbf{1}^T \boldsymbol{f}$. The formal solution to this equation is

$$p_i(t) = \exp\left\{\int_0^t \left[f_i(\boldsymbol{p}(\tau)) - <f>\right]\right\} d\tau\ p_i(0)$$

$$= \exp\left[\int_0^t f_i(\boldsymbol{p}(\tau))\ d\tau\right] p_i(0) \bigg/ \left\{\Sigma_k \exp\left[\int_0^t f_k(\boldsymbol{p}(\tau))\ d\tau\right] p_k(0)\right\},$$

so that the expression for the pragmatic information of the system at time $t$, given its initial conditions is given by

$$\log_2 e \sum_i p_i(t) \left\{\int_0^t \left[f_i(\boldsymbol{p}(\tau)) - <f>\right] d\tau\right\}.$$

This expression can be interpreted as the time averaged growth rate of the system. Since it is well known that some kinds of replicator systems, such as the hypercycle, have limit cycles as their unique long term solution, it must be the case that the pragmatic information for these systems is non-monotonic. We conjecture, but, as yet, have been unable to prove that a necessary and sufficient condition for the pragmatic information for a system to be montonic is that the Jacobian matrix $[\partial f_i/\partial p_j]$ has real eigenvalues.

The time averaged growth rate of the entire biosphere has been far from monotonic, suggesting that the pragmatic information gathered during earlier times is sometimes discarded. There have, for example, been a number of major extinction events, where as many as 90% of the species then existing became extinct. At first glance, this seems to be a powerful argument against intelligent design. However, vestigial organs, such as gills, form and are then reabsorbed during the development of the human embryo, so perhaps extinction events should be viewed as something similar, given the theory of Gaian development.

Another quantity that might indicate Gaian development is the degree of co-evolution in the population dynamics of the biosphere. That co-evolution must play a significant role in the determination of fitness follows from the fact that all predator-prey, parasitic, and symbiotic relationships are co-evolutionary. In fact, it is widely believed that modern eukariotic cells are the result of the ever tighter symbiosis between mitochondria and the rest of the cell.

The theory of pragmatic information suggests a means of quantitatively characterizing the degree of co-evolution between two or more species, namely the pragmatic information in the population of one or more species, given the population of the others. Suppose $\boldsymbol{q}$ is the vector of populations of the "message" species and $\boldsymbol{p}$ is the vector of populations of the "output" species in the replicator equations, then the equations for the growth of $\boldsymbol{p}$ and $\boldsymbol{q}$ are

$$d\boldsymbol{p}/dt = \text{diag}\left[\boldsymbol{p}\right]\left[\boldsymbol{f}(\boldsymbol{p}, \boldsymbol{q}) - \boldsymbol{1} <f(\boldsymbol{p}, \boldsymbol{q})>\right].$$

and

$$d\boldsymbol{q}/dt = \text{diag}\left[\boldsymbol{q}\right]\left[\boldsymbol{g}(\boldsymbol{p}, \boldsymbol{q}) - \boldsymbol{1} <g(\boldsymbol{p}, \boldsymbol{q})>\right],$$

where *g* is the fitness function for the *q*'s. From the point of view of the present theory, co-evolution is indicated when *p* given *q* has significant pragmatic information. Since we are trying to determine whether the dynamics of *q* matter to *p*, we compare *p* as per the recipe of the above equations to *p* computed with *q* "frozen" at its initial value, $q_0$. In other words, we compute not just *p* as per the recipe of the above equations, but also *φ*, the solution to

$$d\varphi/dt = \text{diag}\,[\varphi]\,[f(\varphi, q_0) - 1 < f(\varphi, q_0) >].$$

The pragmatic information, given a specific value for $q_0$, is then

$$\sum_{k,l} p_k(t)\, q_l(t)\, \log_2[p_k(t)\, q_l(t) / \varphi_k(t)\, q_l(0)].$$

We note that *p* and *q* as computed above are *not* independent, because each depends on the history of the vector (*p*, *q*).

Using the formal solution to the replicator equation given above, we have

$$\sum_{k,l} p_k(t)\, q_l(t)\, \left\{\int_0^t \left[f_i(p(\tau), q(\tau)) - 1 < f(p(\tau), q(\tau)) >\right] d\tau + \right.$$

$$\left. \int_0^t \left[g_i(p(\tau), q(\tau)) - 1 < f(p(\tau), q(\tau)) >\right] d\tau - \int_0^t \left[f_i(\varphi(\tau), q_0) - 1 < f(\varphi(\tau), q_0) >\right] d\tau \right\}.$$

The first term in the above expression is the component of the pragmatic information directly involving *p*, as affected by *q*, the last expression as the component directly involving *p*, without information regarding *q*, and the middle term as the component as the indirect influence of *p* on itself, as mediated by *q*. Note that this middle term might not be obvious without the pragmatic information framework. Note also that the relative sizes of these terms allow us to quantify the relative importance of the interaction with the *q* species.

*Randomness in the Message and Computational Models of the Receiver*

We turn now to the question of detecting the temporal connections required to distinguish Gaian development from random mutations. To do this, we must consider the fact that some receivers, because they can apply more sophisticated algorithms to the incoming messages, can make more sophisticated, and thus more "pragmatically informative" decisions. More memory, more different types of primitive calculations, etc. can be used to fashion a better decision making algorithm. However, these quantitative increments obscure the qualitative increases in computing power obtained by ascending the so-called Chomsky hierarchy (Hopcroft, *et. al.*, 2000). The Chomsky hierarchy is a series of abstract "machines" intended to emit a single "accept"/ "reject" output signal if a sequence of input symbols is a member of a particular set of such symbols (the "language"). In our terminology, the machine is making a single "decision," and thus emitting a single symbol, $\omega$. If we are interested in the ongoing response of the observer to a stream of symbols, we consider the machine to be making a series of decisions. We assume, in the most general case, the doubly infinite input sequence of symbols,

$$\alpha_{-n}, \alpha_{-(n-1)}, \alpha_{-(n-2)}, \ldots \alpha_{-1}, \alpha_0, \alpha_1, \ldots \alpha_n, \alpha_{n+1}$$

and that the output process is another doubly infinite, discrete stream of symbols

$$\omega_{-n}, \omega_{-(n-1)}, \omega_{-(n-2)}, \ldots \omega_{-1}, \omega_0, \omega_1, \ldots \omega_n, \omega_{n+1}.$$

We now enumerate the models in the hierarchy and discuss their relevance to the present case.

*The Finite State Machine* [5]

The simplest model in this hierarchy is also the most sophisticated model of a finite computation, the finite state machine (FSM). This model maintains a single state variable, $s$, and a state transition table, $T$, with an entry for each possible input pair $(\alpha_k, s)$. Each such entry, $(\omega_k, s_{next})$, specifies the symbol, $\omega_k$, to output and the next state, $s_{next}$. $\omega_k$ can be assumed to be "no output," without loss of generality.

*The Stack Automaton*

The next more general model in the hierarchy augments the FSM with a "stack," a data structure that stores the state at each iteration according to a "first in, last out" (FIFO) rule. In other words, each time a state is stored, or "pushed" onto the stack, that state, and only that state can be retrieved. Each previous state can only be accessed by "popping," or retrieving the subsequently stored states from the stack, in reverse order of storage. Once retrieved, a state is "forgotten" if it is overwritten in the FSM by a subsequent state. The stack automaton can clearly implement an FSM by simply ignoring the stack, demonstrating that the former is strictly more powerful than the latter.

In theory, the stack can grow arbitrarily long; in practice, stack overflow is sometimes an issue for most problems, but less so as memory sizes increase. However, the amount of data that might need to be analyzed to identify departures from randomness in mutation rates could easily exceed the largest of currently available computer memories. It seems, for example, that mutation rates for a given gene can depend on the three dimensional structure of the proteins for which the gene codes (Liò and Goldman, 1998).

---

[5] Technically, a finite state machine with output is known as a Mealy machine. We retain the more familiar terminology.

*Turing Machine*

A Turing machine, which sits atop the Chomsky hierarchy, is an FSM with the following modifications and additions:
- a doubly infinite tape of discrete cells, each of which can store a single value,
- a mechanism for reading from and/or writing to a single cell, and, possibly, advancing the tape one cell to the left or right of the current one,
- an enlargement of the current state of the FSM to include, not just the current internal state, but also the contents of the cell currently under the read/write mechanism,
- an enlargement of the next state of the FSM to include the contents, if any, to be written to the cell currently under the read/write mechanism.

Thus, the state transition table becomes a mapping from the current state of the FSM and the data on the tape to the next state of the FSM

The Turing machine is widely considered to be the most powerful of the abstract computing paradigms, as Turing machines can implement every other known computing paradigm. For many purposes, the assumption of ain infinite tape is reasonable because the amount of memory actually available in modern computers is typically much larger than the data that is being processed. This fact has made the Turing machine an attractive paradigm for theoretical studies. However, the practical problems surrounding the stack of the stack automaton arise even more forcefully for the doubly infinite tape of the Turing machine.

All three of the above computing paradigms can be analyzed using the same mathematical framework, but the simplest case involves the FSM. We therefore consider that case first, and introduce some notation. We adopt the convention that a superscript identifies a particular value for a symbol or a particular state, and a subscript identifies the location of the symbol in the input sequence or the resulting state. Thus, we let $\alpha^l$ be the $l^{th}$ of the $L$ possible symbols that can appear at a given position in the message, $\alpha_n$ be the symbol that appears in the $n^{th}$ position in the message (whatever $\alpha^l$ that symbol happens to be), and $\boldsymbol{\alpha}_n$ be the sequence of message symbols up to and including the $n^{th}$ (so that $\boldsymbol{\alpha}_n = \boldsymbol{\alpha}_{n-1} \alpha_n$). We assume further that there are $F$ states, $S^1, S^2, \ldots, S^F$, in the FSM and that the indicator function $\mathbf{1}\{(S^i, \alpha^l) \rightarrow (S^j, \alpha^m)\} = 1$ if and only if symbol $\alpha^l$ can induce a transition between state $S^i$ and state $S^j$ and if the symbol $\alpha^m$ follows $\alpha^l$ in the input sequence.

The dynamics of the FSM, and thus its sequence of output $\omega$'s, is determined by the transition matrix $\boldsymbol{T}$, whose $p, q^{th}$ component is given by

$$T_{pq} = Pr\{\alpha_n = \alpha^l \parallel \boldsymbol{\alpha}_{n-1}\} \; \mathbf{1}\{(S^i, \alpha^l) \rightarrow (S^j, \alpha^m)\},$$

where $Pr\{α_n = α^l \| \mathbf{α}_{n-1}\}$ is the probability that $α^l$ appears immediately after $\mathbf{α}_{n-1}$ and $p = (i - 1) L + l$ and $q = (j - 1) L + m$. Consider first the special case in which

$$Pr\{α_n = α^l \| \mathbf{α}_{n-1}\} = Pr\{α_n = α^l | α_{n-1} = α^k \};$$

in other words, the message is a stationary[6], first order Markov process. $T$ is then the $LF \times LF$ matrix whose components are given by $T_{pq} = Pr\{α_n = α^l | α_{n-1} = α^k \} \mathbf{1}\{(S^i, α^l) \rightarrow (S^j, α^m)\}$. We also introduce the $LF$ dimensional vector $\mathbf{v}_{(n)} = (π_n^1, π_n^2, ..., π_n^F, π_n^1, π_n^2, ..., π_n^F, ..., π_n^1, π_n^2, ..., π_n^F)$, which has $L$ copies of the vector of probabilities that the FSM is in each of its $F$ states after input $α_n$. Since the matrix-vector product of $\mathbf{v}_{(n-1)} T$ is $\mathbf{v}_{(n)}$, we see that the dynamics are those of a stationary Markov chain with states corresponding to the $LF$ possible ordered pairs $(S^i, α^l)$. While the detailed dynamics depends on the details of the $T$ matrix, the theory of finite Markov chains (Isaacson and Madsen, 1976) tells us that, generically, the system will spend a finite amount of time in a set of so-called transient states, after which it settles into a set of ergodic states. Once there, as successive symbols in the message are processed, the probability distribution of states converges exponentially to an invariant distribution that is independent of the initial distribution. Further investigation is required, but it appears that the long term dependencies on mutations required for Gaian development cannot occur in this case.

Assuming that the probabilities of successive message symbols depend on $M > 1$ previous symbols does not qualitatively change things, because $T$ has the finite dimension $FL^M$. The situation is, however, qualitatively different if the Markov chain is "sufficiently" non-stationary in a sense that is made precise in Isaacson and Madsen, 1976. This situation and the situation in which the symbol sequence is non-Markovian must be left to future research. A complication that arises in this latter case is the fact that the $T$ matrix becomes infinite dimensional.

### *Summary and Thoughts about Future Research*

In contrast to standard evolutionary theory, Gaian development emphasizes co-evolutionary processes, coordinated genetic mutations, and other mechanisms for transmitting information from one evolving species to another. This paper is a first attempt to analyze these information flows, using results from the quantitative theory of pragmatic information summarized above. However, this theory already suggests ways of quantifying

- the rate of evolution

- the decomposition of that rate into a co-evolutionary component and a component dependent only on the single species under consideration

---

[6] This stationarity assumption is a standard one in current models of gene substitution (See Liò and Goldman, 1998, for details.).

- the rate at which an observer with a finite memory would lose the ability to "remember" the distant past

In future work, we hope to obtain corresponding results for an observer with effectively infinite memory, which may be qualitatively different.

While much of the advance of the above theory probably will come from a closer contact with empirical results, pragmatic information might also be of use in understanding the instabilities that can arise as the result of network effects. These effects might underlie such diverse phenomena as the mass extinctions of biological species (Solé *et. al.*, 1999) and stock market crashes (Sornette, 2003).

We close by making the distinction between Gaian development and that stalking horse for Creationism, "Intelligent Design." The latter is the view that the biosphere was shaped by an intelligent agency outside of nature, a view that has been roundly rejected by the scientific community. In contrast, this paper argues for a naturalistic extension to the standard Darwinian paradigm that may Gaian development, a version of intelligent design in which the intelligence is an emergent property of the biosphere, is more scientifically plausible.

*Acknowledgements*

This work was partially supported by the estate of David P. Weinberger, Esq., the author's uncle.

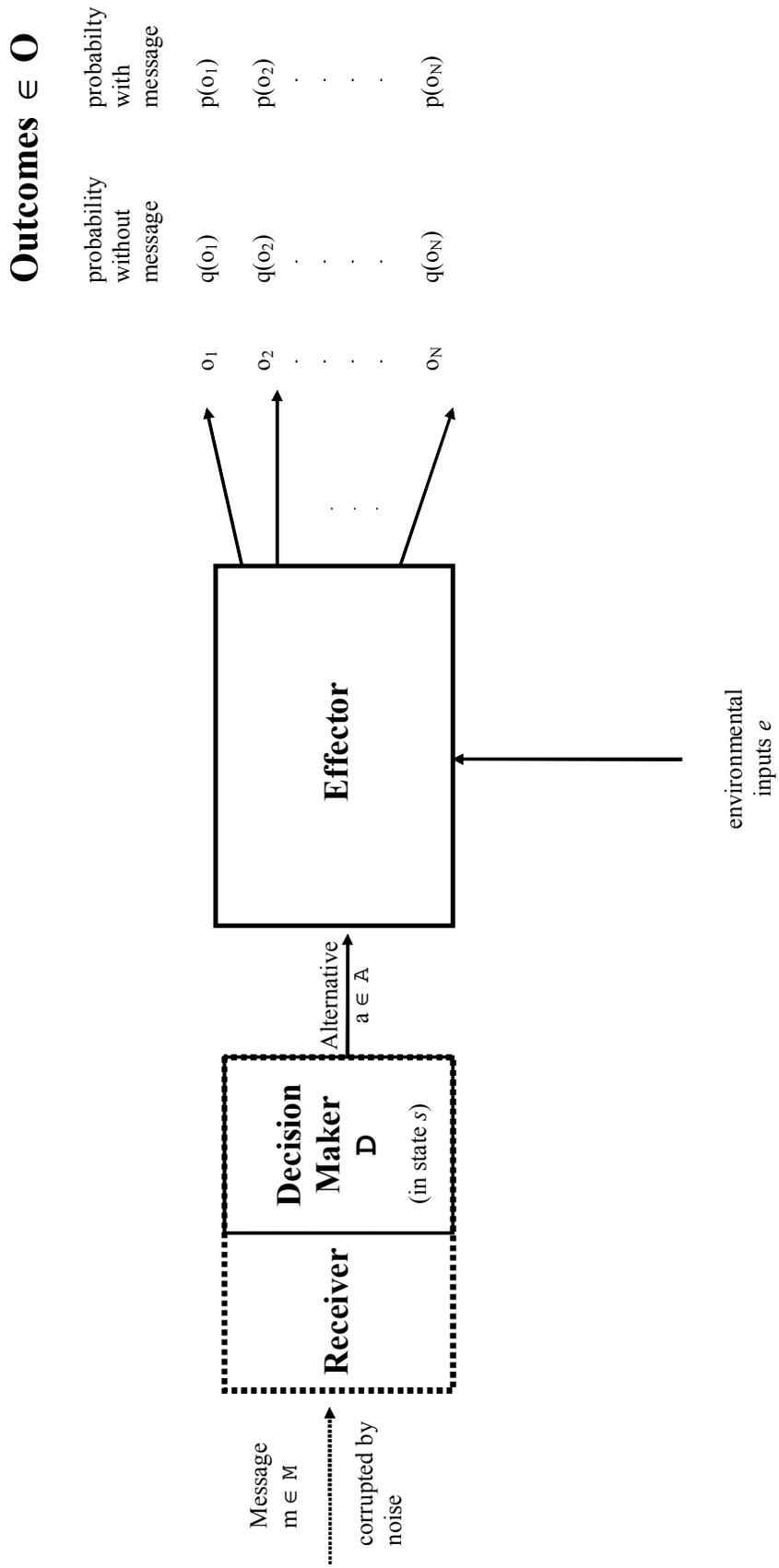

Figure 1